# Algorithmic labeling in hierarchical classifications of publications: Evaluation of bibliographic fields and term weighting approaches

Peter Sjögårde[a,b,x], Per Ahlgren[c,y], Ludo Waltman[d,z]

[a]Health Informatics Centre, Department of Learning, Informatics, Management and Ethics, Karolinska Institutet, Stockholm, Sweden
[b]University Library, Karolinska Institutet, Stockholm, Sweden
[c]Department of Statistics, Uppsala University, Uppsala, Sweden
[d]Centre for Science and Technology Studies, Leiden University, Leiden, The Netherlands

ORCID:
[x]https://orcid.org/0000-0003-4442-1360
[y]https://orcid.org/0000-0003-0229-3073
[z]https://orcid.org/0000-0001-8249-1752

Email: peter.sjogarde@ki.se; per.ahlgren@uadm.uu.se; waltmanlr@cwts.leidenuniv.nl

Corresponding author: Peter Sjögårde, University Library, Karolinska Institutet, 17177 Stockholm, Sweden

**Abstract**
Algorithmic classifications of research publications can be used to study many different aspects of the science system, such as the organization of science into fields, the growth of fields, interdisciplinarity, and emerging topics. How to label the classes in these classifications is a problem that has not been thoroughly addressed in the literature.

In this study we evaluate different approaches to label the classes in algorithmically constructed classifications of research publications. We focus on two important choices: the choice of (1) different bibliographic fields and (2) different approaches to weight the relevance of terms.

To evaluate the different choices, we created two baselines: one based on the Medical Subject Headings in MEDLINE and another based on the Science-Metrix journal classification. We tested to what extent different approaches yield the desired labels for the classes in the two baselines.

Based on our results we recommend extracting terms from titles and keywords to label classes at high levels of granularity (e.g. topics). At low levels of granularity (e.g. disciplines) we recommend extracting terms from journal names and author addresses. We recommend the use of a new approach, term frequency to specificity ratio, to calculate the relevance of terms.

# Introduction

In recent years scientometricians have developed methods for algorithmically constructing classifications of research publications based on relations between individual publications. This has been done using large publication sets of tens of millions of publications (Boyack & Klavans, 2014; Sjögårde & Ahlgren, 2018; Waltman & van Eck, 2012). The obtained classifications have been used for various applications, such as identification of research topics and specialties, normalization of citations, measuring interdisciplinarity and mapping research fields (Ahlgren et al., 2018; Milanez et al., 2016; Ruiz-Castillo & Waltman, 2015; Sjögårde & Ahlgren, 2020; Small et al., 2014; Šubelj et al., 2016; Wang & Ahlgren, 2018). In order to be useful for a wide range of applications, the classes in classifications obtained by algorithmic methods must have labels.[1] Labels make it possible for users to get a clue about the contents of each class. Hierarchical classifications with labeled classes make it possible for users to browse large document collections (Cutting et al., 1992; Seifert et al., 2014). Perianes-Rodriquez and Ruiz-Castillo (2017) point out that the applicability of publication-level classifications is limited at higher granularity levels because of the labeling problem. Several conceptual and methodological challenges related to algorithmically constructed publication-level classifications (ACPLCs) have already received considerable attention in the literature (Ahlgren et al., 2020; Boyack & Klavans, 2020; Klavans & Boyack, 2017; Sjögårde & Ahlgren, 2018, 2020; Šubelj et al., 2016; Velden, Boyack, et al., 2017; Waltman et al., 2020), but the problem of labeling the classes in these classifications has not been much discussed. This problem is the focus of our paper.

Classifications of research publications have some special features. Bibliographic databases such as Web of Science, Scopus and Dimensions contain information about journal articles and are more structured than most other sources commonly used in the algorithmic clustering literature, for example web pages, blog posts, Wikipedia pages or text documents (Carmel et al., 2009; Hennig et al., 2014; S.-T. Li & Tsai, 2010). In hierarchical classifications the subject orientation of classes needs to be determined at different semantic levels. In their publications researchers express the subject content of their work through the terminology they use. In addition to the full text of a publication, the subject orientation is expressed in titles, abstracts and keywords — information available in bibliographic databases. The terminology used in these bibliographic fields usually expresses the subject orientation at a detailed level. It is less likely that the subject orientation is expressed at higher semantic levels, for example by using terms such as "mathematics", "physiology" or "pedagogics". When labeling classes in hierarchical classifications there is a need to express the subject orientation both at detailed levels, for example at the level of topics, and at higher semantic levels, for example at the level of disciplines. Terminology that corresponds to higher semantic levels may be found in journal names, for example "Scientometrics", or in addresses, for example "Department of Neuroscience". In this paper we investigate if this information can indeed be used for class labeling.

Different methods for calculating a term's relevance for labeling a particular class have been proposed, for example Chi-square, term frequency-inverted document frequency (TF-IDF) and Jensen-Shannon Divergence (JSD). Typically such approaches take into account the frequency of the term in the class and the so-called "discriminating power" of the term (Manning et al., 2008). A term with a high discriminating power describes the subject orientation of a class in a way that enables the class to be distinguished from other classes. Terms with a high discriminating power are specific to a particular class. In this paper we therefore say that these terms have a high specificity. Terms that combine a high frequency with a high specificity can be expected to be most suitable for class labeling. To our knowledge, there is no study in the field of scientometrics that evaluates the use of different approaches for determining the relevance of terms for labeling classes in ACPLCs.

---

[1] Since our goal is to create a classification, we use the term "class" to denote a set of publications, where the set is obtained using a clustering methodology. The terms "cluster" and "community" are also regularly used in the literature to denote such a set of publications.

In this paper, we address the problem of labeling classes of research publications obtained using algorithmic methods. Labels that make the subject orientation of classes easily interpretable for users improve the usefulness of such classes. We restrict the study to two aspects of class labeling: the choice of (1) different bibliographic fields and (2) different approaches to weight the relevance of terms. Other aspects may also be of relevance for class labeling, such as the use of different term extraction approaches. However, we leave these aspects to future studies. Furthermore, we focus on the particular context of labeling classes of large sets of research publications and we do not consider other kinds of documents. We perform the evaluation using two baselines: the Medical Subject Headings (MeSH) in MEDLINE and the Science-Metrix journal classification (SMJC).

**Related research**

In this section, we discuss literature that is of relevance for the problem of labeling classes (or clusters) in hierarchical ACPLCs. We do not consider the broader literature on related issues such as topic modelling (see e.g. Aggarwal & Zhai, 2012; Blei & Lafferty, 2007; Blei & Lafferty, 2009; Suominen & Toivanen, 2016) and the calculation of document relevance to search queries (see e.g. Carmel et al., 2006; Deerwester et al., 1990; Salton & Buckley, 1988; Spärck Jones, 2004). We also need to point out that the problem of class labeling of ACPLCs is different from the task of text classification. We are not bound to a pre-existing taxonomy, for example MeSH or the Library of Congress subject headings. On the contrary, the classes of the classification are created by a clustering methodology and labeled afterwards. Therefore, we do not discuss text classification approaches. Nevertheless, we recognize that text classification approaches can be an alternative to construct classifications of research publications.

As phrased by Treeratpituk and Callan (2006), approaches that aim to label classes in a hierarchical classification are based on the hypothesis that "by comparing the word distribution from different parts of the hierarchy, it should be possible to assign appropriate labels to each cluster in the hierarchy." They performed relevance ranking of single words, bigrams and trigrams using an algorithm that takes into account the normalized term frequency (proportion of documents containing a term) in both the class of interest and the parent class. The algorithm also takes term length into account, preferring long terms over short ones. An interesting feature of their approach is that they use training data to determine a cut off for the number of terms to be chosen as labels.

Carmel et. al (2009) evaluated different approaches for labeling classes in (1) a classification of news items and (2) a classification of web pages. They evaluated the different approaches by trying to predict labels assigned manually to classes. An approach using Wikipedia information to obtain labels was proposed by the authors. This approach was compared to JSD, Chi-square, Mutual Information and CTF-CDF-IDF. In the experiments, the Wikipedia approach performed best followed by the JSD approach.

Muhr et al. (2010) compared several term weighting approaches for labeling of classes using three different data sets derived from the Open Directory Project, Wikipedia and TREC Ohsumed. To each of the approaches they added a component that takes into account the distribution of a term in a class' child classes. The idea of this approach is to favor terms that are equally likely to be represented in child classes over terms that are represented in just one or a few of the child classes. The authors concluded that the performance of the approaches is enhanced by taking this hierarchical factor into account. They proposed a new term weighting approach which they call ICWL (Inverse Cluster frequency Weighted Labeling). This approach performed best, followed by JSD.

Also Mao et al. (2012) compared several term weighting approaches for labeling of clusters, in their case using two documents sets, one derived from Yahoo! Answers and another one from Wikipedia. Like Muhr et al. (2010) they added a component that takes into account the distribution of a term in a class' child classes. They concluded that taking hierarchical factors into account enhances the performance of the labeling approach.

There are some studies in the field of scientometrics that are relevant to our work. A special issue on topic extraction from collections of research publications was published in *Scientometrics* in 2017. The various contributions used different approaches to extract topics from a publication set in the field of astrophysics (for a summary of the results see Velden, Boyack, et al., 2017). Some of the articles in the special issue focused on labeling of classes. Velden, Boyack, et al. (2017) used terms from a thesaurus to label clusters and weighted the extracted terms using Normalized Mutual Information (NMI). NMI was also used by Koopman and Wang (2017), who extracted terms from titles and abstracts to label classes. Velden, Yan, et al. (2017) used "journal signatures" to label classes at a higher sematic level.

A simple approach that combines term frequency and term specificity to label classes was proposed by Waltman and van Eck (2012). In this approach, term relevance is calculated by dividing a term's frequency in a class by the sum of the frequency of the term in the parent class and a parameter value (set to 25 in the particular case). The parameter value can be adjusted to give more or less weight to either term frequency or term specificity.

Several approaches have also been proposed to create labels using external sources such as Wikipedia, ontologies and thesauri (Allahyari & Kochut, 2015; Carmel et al., 2009; Hotho et al., 2003; Velden, Boyack, et al., 2017; Velden, Yan, et al., 2017).

Sophisticated methods have been proposed to extract terms from a text corpus and to use these terms to create class labels. Li et al. (2015) used dependency parsing along with a set of rules to extract terms. This method is able to identify candidate labels in which two nouns are separated by a preposition, for example "degrees of freedom". Also Word2vec can be used to retrieve phrases from a text corpus (Mikolov et al., 2013). Word2vec transforms a text corpus to numerical vectors. It provides n-grams that can be used as candidate labels.

Previous studies have not made clear which bibliographic information is most suited to create labels in hierarchical ACPLCs. Neither has previous research evaluated different approaches to weight and rank the relevance of terms for labeling classes in ACPLCs. In this paper, we aim to address these two issues.

**Study design**

We use two baseline classifications, one based on MeSH and one based on SMJC, to evaluate two key aspects of different labeling approaches: the choice of (1) different bibliographic fields and (2) different approaches to weight the relevance of terms.

To evaluate a labeling approach using MeSH as a baseline, each MeSH term is treated as a class. The MeSH term is considered as the label of the class and the publications to which the MeSH term has been assigned as major descriptor are considered as the publications belonging to the class. A labeling procedure extracts terms from bibliographic fields (other than the MeSH field) and ranks the extracted terms based on their estimated relevance for labeling a class. For a given MeSH term, a labeling procedure is successful if the MeSH term is included among the top $N$ most relevant terms identified by the labeling procedure. The evaluation methodology is similar for the SMJC baseline. In this case we treat the journal categories as the baseline classes. All publications in the journals in a journal category are considered to belong to the same class. The name of the journal category is considered as the label of the class. For a given SMJC category, a labeling procedure is successful if the name of the category is included among the top $N$ most relevant terms identified by the labeling procedure.

MeSH is a controlled vocabulary provided by the U.S. National Library of Medicine. It is used for subject indexing of primarily research articles and contains terms at different semantic levels, with a depth of at most 13 levels.[2] Before 2001 MeSH indexing was done manually and from 2001 manual indexing has been supported by computer generated recommendations. In 2011 algorithmic MeSH indexing was introduced for a small set of journals. The number of journals for which algorithmic MeSH indexing is used has increased since then (*NLM Medical Text Indexer (MTI)*, n.d.). For our analysis we use publications from the years 2006-2010, so we do not use publications with algorithmically assigned MeSH terms. The same MeSH term can be located at multiple places in the MeSH tree. We disregard MeSH terms for which this is the case. We also consider only MeSH terms that consist of exactly one noun phrase (see the section "Term extraction"). In the SMJC baseline, classes may have labels consisting of multiple noun phrases, for example "Nanoscience & Nanotechnology". In this case, a successful approach ranks either "Nanoscience" or "Nanotechnology" among the top *N* most relevant terms for the class (see the section "Evaluation methodology" for further details).

MeSH provides a baseline describing the subject orientation of biomedical and health related literature primarily at a detailed level. It has been created to enable retrieval of comprehensive search results from search terms. For all publications to which a particular term has been assigned, an indexer has considered the term to be of relevance. We therefore consider the term to be a suitable label for this set of publications.

The vocabulary of SMJC is of a different nature than the MeSH vocabulary. Archambault et al. (2011) discuss how they created SMJC. Traditional journal classifications were used to create the categories in SMJC. Journals were algorithmically assigned to these categories and some manual changes were made thereafter. The classification has three levels. We use the most granular level, which consists of 176 classes, and the middle level, which consists of 22 classes. By using SMJC as a baseline we are able to evaluate term labeling at higher sematic levels. This is important for labeling classes at higher levels in hierarchical classifications.

We recognize that neither MeSH nor SMJC constitutes a perfect ground truth. However, we use the baselines to compare different approaches and their relative performance.

## Data

**Bibliographic fields**

We use the Karolinska Institutet (KI) in-house version of Web of Science for the analyses.[3] Web of Science is provided by Clarivate Analytics and the KI system contains data from the Science Citation Index Expanded, Social Sciences Citation Index and Arts & Humanities Citation Index. We base the analysis on the publication types "Article" and "Review". We restrict the use of bibliographic fields to fields that are independent of the choice of database and that can be expected to contain subject-related information. Furthermore, only information that can be obtained from publications without pre-processing is used, with the exception of address information, which has been pre-processed by Clarivate Analytics to some extent. We make use of titles, abstracts, author keywords (hereafter we refer to author keywords simply as "keywords"), journal names and the suborganization field in the author addresses (hereafter we refer to the suborganization field simply as the "address").[4] We do not use database specific information such as the Web of Science journal categories, MeSH terms (which in this case would not be independent of the MeSH baseline) or any thesaurus provided by a database provider.

---

[2] For a brief description of how the indexing is performed, see "Principles of MEDLINE Subject Indexing" at https://www.nlm.nih.gov/bsd/disted/meshtutorial/principlesofmedlinesubjectindexing/principles/index.html [2020-09-03]

[3] Certain data included herein are derived from the Web of Science ® prepared by Clarivate Analytics ®, Inc. (Clarivate®), Philadelphia, Pennsylvania, USA: © Copyright Clarivate Analytics Group ® 2020. All rights reserved.

We expect terms extracted from titles and keywords to be more likely to perform well at lower levels of aggregation, but to be less suited to label classes at higher levels in hierarchical classifications. On the other hand we expect journal names and addresses to perform better at higher, less granular levels than at lower, more granular levels. Table 1 shows an example of a publication record in which terms that provide information about the subject orientation have been underlined. This record includes specific terms such as "compositional gradient polymeric films" in the title and "water vapor permeability" in the list of keywords, and some broader terms such as "polymers" and "molecular engineering" in the journal and address fields.

*Table 1: Example of a publication record in PubMed/MEDLINE (PMID: 30966711). Terms that may provide information about the subject orientation of the publication are underlined.*

| **Title** | Preparation of Compositional Gradient Polymeric Films Based on Gradient Mesh Template |
|---|---|
| **Keywords** | compositional gradient; filling method; gradient mesh template; hydrophilic/hydrophobic; water vapor permeability |
| **Journal** | Polymers |
| **Addresses** | Shandong Provincial Key Laboratory of Molecular Engineering, School of Chemistry and Pharmaceutical Engineering, Qilu University of Technology (Shandong Academy of Sciences), Jinan 250353, China. College of Chemistry, Chemical Engineering and Materials Science, Shandong Normal University, Jinan 250100, China. |

Address information in Web of Science has been pre-processed by Clarivate Analytics. As part of this process, addresses have been abbreviated. We deabbreviate addresses and remove some stop words that are not meaningful as subject descriptors, such as "department", "school" and "faculty". This process is quite complex and we therefore describe it in detail in supplementary material.

For the MeSH baseline, five bibliographic fields or combinations of bibliographic fields are considered: (1) titles, (2) keywords, (3) titles and keywords, (4) titles, keywords and abstracts, and (5) titles, keywords, journals and addresses.

For the SMJC base line the following fields are considered: (1) journals, (2) addresses, (3) journals and addresses, (4) titles and keywords, and (5) journals, addresses, titles and keywords.

**Term extraction**

To extract terms from bibliographic fields we use the same approach as Waltman and van Eck (2012). This approach is based on the idea that "[m]ost terms have the syntactic form of a noun phrase" (Justeson & Katz, 1995; Kageura & Umino, 1996). We operationalize noun phrases as sequences of adjectives and nouns that end with a noun. We use the Stanford Core NLP software (Manning et al., 2014).[5] The part-of-speech tagger is used to identify sequences of adjectives and nouns (Toutanova et al., 2003; Toutanova & Manning, 2000). To unify singular and plural words, we lemmatize each word. We also replace hyphens with spaces, convert all text to lower case and keep only alphanumeric characters and spaces. As an illustration, Table 2 shows the noun phrases extracted from the publication record presented in Table 1.

---

[4] The suborganization field is provided in the raw XML data of Web of Science. It is also searchable in the advanced search in the Web of Science web interface. The suborganization field typically refers to an internal department within an organization, for example an institute within a university. In the following example the bold formatting highlights the suborganization field: "KTH Royal Inst Technol, **Sch Educ & Commun Engn Sci ECE**, Stockholm, Sweden".

[5] Stanford CoreNLP is available at https://stanfordnlp.github.io/CoreNLP/.

In the remainder of this paper, we will refer to the extracted noun phrases as terms.

*Table 2: Noun phrases extracted from the publication record presented in Table 1.*

| Bibliographic field | Extracted noun phrase |
|---|---|
| Title | preparation |
| Title | compositional gradient polymeric film |
| Title | gradient mesh template |
| Journal | polymer |
| Address | science |
| Address | molecular engineering |
| Address | shandong provincial key |
| Address | pharmaceutical engineering |
| Address | chemistry |
| Address | materials science |
| Address | chemical chemical engineering |
| Keyword | water vapor permeability |
| Keyword | method |
| Keyword | gradient mesh template |
| Keyword | compositional gradient |
| Keyword | hydrophobic |
| Keyword | hydrophilic |

**Creation of MeSH baseline**

The KI bibliometric database contained 29,638 MeSH terms at the time of data extraction (January 2020). After restricting the analysis to MeSH terms that correspond to exactly one noun phrase, 22,642 terms remained. So 77% of the MeSH terms were included in this study.

We further restricted our analysis to MeSH terms that have exactly one location in the MeSH tree. We refer to the set of publications to which a given MeSH term is assigned as a MeSH class. Only MeSH classes containing at least 50 publications were considered. For each of these MeSH classes, noun phrases were extracted from the bibliographic fields of the publications belonging to the class. For efficiency reasons noun phrases occurring in fewer than three publications in a MeSH class were not taken into account. MeSH classes with no noun phrases that occur in at least three publications were disregarded. After applying these restrictions, the MeSH baseline contained 5,209 classes covering about 2.5 million unique publications. Figure 1 shows the number of classes per level in the MeSH tree. Higher levels contain smaller and more fine-grained classes. We omitted the root level with 16 broad classes from the analysis. Summary statistics on the number of publications per MeSH class are shown in Table 3. Table 4 shows summary statistics on the number of terms per MeSH class.

*Figure 1: Number of MeSH classes included in the analysis per level in the MeSH tree.*

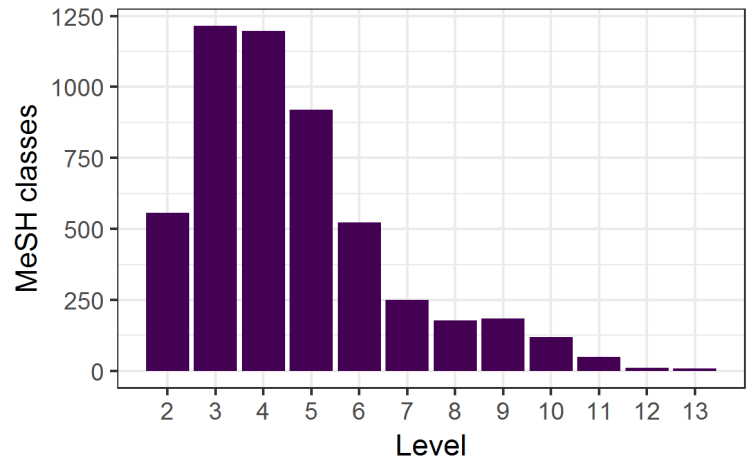

*Table 3: Summary statistics on the number of publications per MeSH class.*

| Level | # Classes | Mean | Min | Q1 | Median | Q3 | Max |
|---|---|---|---|---|---|---|---|
| 2 | 557 | 9,559.3 | 50 | 371 | 1,949 | 8,275 | 138,059 |
| 3 | 1,215 | 4,046.7 | 50 | 204 | 764 | 2,666 | 188,688 |
| 4 | 1,197 | 1,963.8 | 50 | 151 | 412 | 1,307 | 82,955 |
| 5 | 919 | 1,404.0 | 50 | 117 | 297 | 823 | 82,892 |
| 6 | 522 | 1,124.4 | 50 | 112 | 230 | 673 | 79,864 |
| 7 | 249 | 1,513.6 | 51 | 115 | 284 | 747 | 79,121 |
| 8 | 178 | 1,037.6 | 50 | 95 | 185 | 641 | 24,672 |
| 9 | 185 | 653.4 | 50 | 81 | 158 | 429 | 14,364 |
| 10 | 120 | 433.3 | 50 | 76 | 117 | 319 | 7,568 |
| 11 | 48 | 308.5 | 50 | 71 | 105 | 189 | 4,333 |
| 12 | 11 | 652.3 | 74 | 140 | 219 | 346 | 3,859 |
| 13 | 8 | 572.0 | 52 | 74 | 517 | 901 | 1,340 |

*Table 4: Summary statistics on the number of terms per MeSH class.*

| Bibliographic field(s) | Mean | Min. | Q1 | Median | Q3 | Max. |
|---|---|---|---|---|---|---|
| Titles | 422.3 | 1 | 14 | 50 | 220 | 26,678 |
| Keywords | 396.2 | 1 | 13 | 45 | 198 | 21,241 |
| Titles and keywords | 732.5 | 1 | 28 | 94 | 398 | 40,419 |
| Titles, keywords and abstracts | 3,761.8 | 18 | 218 | 651 | 2,293 | 194,787 |
| Titles, keywords, journals and addresses | 1,159.9 | 8 | 66 | 200 | 723 | 55,628 |

**Creation of SMJC baseline**

The SMJC baseline consists of 198 classes, 176 at level 3 and 22 at level 2. These classes cover 4.9 million unique publications. At a given level a publication can belong to only one class. Like in the case of the MeSH baseline, classes at the top level were disregarded. Summary statistics on the number of publications per SMJC class are shown in Table 5. Table 6 shows summary statistics on the number of terms per SMJC class.

*Table 5: Summary statistics on the number of publications per SMJC class.*

| Level | # Classes | Mean | Min | Q1 | Median | Q3 | Max |
|---|---|---|---|---|---|---|---|
| 2 | 22 | 221,616.7 | 4948 | 25,545 | 139,850 | 271,541 | 1,244,094 |
| 3 | 176 | 27,702.1 | 228 | 7,064 | 17,144 | 39,315 | 172,164 |

*Table 6: Summary statistics on the number of terms per SMJC class.*

| Bibliographic field(s) | Mean | Min. | Q1 | Median | Q3 | Max. |
|---|---|---|---|---|---|---|
| Journals | 110.2 | 4 | 36 | 65 | 109 | 2,013 |
| Addresses | 2,813.7 | 4 | 496 | 1,493 | 3,249 | 50,566 |
| Journals and addresses | 2,891.8 | 7 | 523 | 1,568 | 3,316 | 51,780 |
| Titles and Keywords | 10,029.5 | 10 | 1,753 | 5,046 | 10,843 | 187,435 |
| Journals, addresses, titles and keywords | 12,698.3 | 17 | 2,334 | 6,494 | 13,674 | 234,687 |

## Methods

### Term weighting approaches

Several approaches can be used to rank the terms in a class. Approaches usually combine two different aspects: the frequency of a term in the class and the specificity of the term. The specificity can be approximated by comparing the frequency of a term in a class with the frequency of the term in a reference set. It is possible to use different reference sets in this calculation. One approach is to compare the term frequency in a class with the term frequency in the *whole* collection of publications in the database. This approach fails to take the context of a term into account. For example, consider the term "humanities", which in general is a very broad term. Suppose that "humanities" occurs a lot in a particular class and that the parent class of this class is about scientometrics. In this particular context, "humanities" should have a rather high specificity. The term "humanities" indicates that the focal class must be at least partly about scientometric methods applied in the humanities. However, if a weighting approach uses the whole collection of publications in the database as the reference set, "humanities" will have a low specificity, because "humanities" also occurs a lot in the whole publication collection. On the other hand, in scientometrics "humanities" is not used very frequently. Hence, if the *parent class* is used as the reference set, "humanities" will have a higher specificity. For this reason, we choose to use the parent class as the reference set in all weighting approaches that we consider.

Figure 2 shows the terms occurring in at least three publications for an example class. The *y*-axis shows the frequency of a term. The *x*-axis shows the specificity of a term, calculated as the number of occurrences of the term in the example class as a share of the number of occurrences of the term in the parent class. Ideally a term is located in the top-right corner in Figure 2, indicating that the term has both a high frequency and a high specificity. However, as the figure illustrates, there is a trade-off between frequency and specificity — if more frequent terms are preferred, the specificity of the terms tends to be low, and if more specific terms are preferred, the frequency tends to be low.

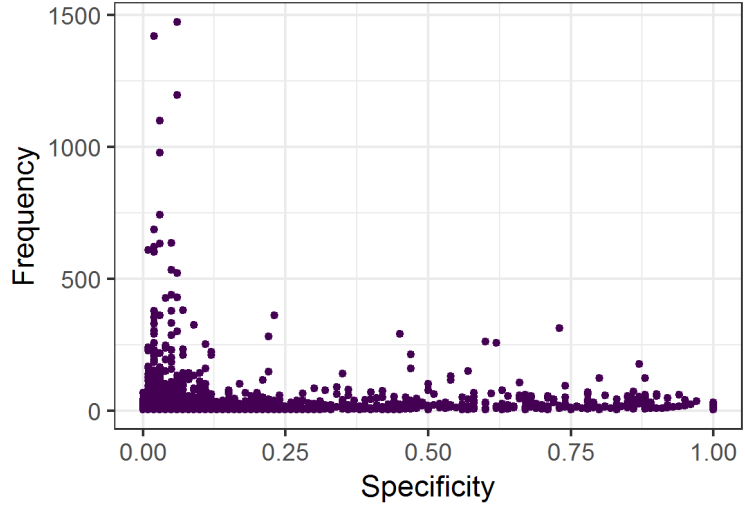

*Figure 2: Relation between term frequency and term specificity for an example class.*

Let $D$ be a set of publications and $C = \{c_1,...,c_k\}$ a set of classes into which the publications in $D$ have been grouped. Since we are dealing with clustering solutions at different hierarchical levels, classes in $C$ at different levels can be overlapping. At a given level, classes are pairwise disjoint in the case of the SMJC baseline. In other words, two classes at the same level do not have any common publications. However, classes at the same level can be overlapping in the case of the MeSH baseline.

Let $T = \{t_1,...,t_n\}$ be the set terms that occur in the publications in $D$. We want to rank these terms, for a given class $c_j$ in $C$, according to their relevance for labeling $c_j$. Furthermore, let $c_p$ be the parent class of $c_j$ and $A(t_i,c_j)$ the relevance of $t_i$ to $c_j$ according to the term weighting approach $A \in \{\text{Chi-square,JSD,JSDQ,TF-IDF,WvE,TFS}\}$ (the approaches are described below).

Finally, we note that we only take into account whether a term occurs in a publication or not. We do not take into account the number of times a term occurs in a publication. Therefore, the frequency of $t_i$ in $c_j$, denoted by $tf(t_i,c_j)$, is the number of publications in $c_j$ in which $t_i$ occurs.

*Chi-square*

Chi-square compares an observed distribution with an expected distribution. In our case, the distribution of the terms in $T$ over the classes in $C$ is compared to the corresponding expected distribution. However, since we are not interested in the whole distribution, we only use one element of the calculation of Chi-square, namely the element in which the frequency of $t_i$ in $c_j$ is compared to the corresponding expected frequency. Chi-square is therefore given by

$$Chi\text{-}square(t_i,c_j) = \begin{cases} \dfrac{(tf(t_i,c_j) - expected)^2}{expected} & \text{if } tf(t_i,c_j) - expected > 0 \\ 0 & \text{if } tf(t_i,c_j) - expected \leq 0 \end{cases} \quad (1)$$

where the expected frequency equals

$$expected = tf(t_i, c_p) \times \frac{|c_j|}{|c_p|}$$

Normally Chi-square would measure the extent to which the frequency of $t_i$ in $c_j$ deviates from the expected frequency regardless of whether $t_i$ occurs more or less frequently than expected. We have added the condition that the frequency of $t_i$ in $c_j$ should be higher than expected. If this is not the case, the value of Chi-square is set to zero.

### *Jensen-Shannon Divergence (JSD)*

Like Chi-square, JSD compares two distributions. We start from the definition of JSD provided by Muhr (2010). According to this definition, JSD is given by

$$JSD(t_i, c_j) = P(t_i) \log \frac{P(t_i)}{M(t_i)} + Q(t_i) \log \frac{Q(t_i)}{M(t_i)} \quad (2)$$

$$P(t_i) = \frac{tf(t_i, c_{\text{ref}})}{\sum_{t \in T} tf(t, c_{\text{ref}})}; \quad Q(t_i) = \frac{tf(t_i, c_j)}{\sum_{t \in T} tf(t, c_j)}; \quad M(t_i) = \tfrac{1}{2}(P(t_i) + Q(t_i))$$

where $c_{\text{ref}} = c_p \setminus c_j$ is the *reference collection* for $c_j$, defined as the publications in $c_p$ with the exclusion of the publications in $c_j$.

JSD has an important problem. A term may be highly ranked by JSD not because it occurs a lot in $c_j$ but because it occurs a lot in $c_{\text{ref}}$. The example below presents a hypothetical case demonstrating this problem.

**Example**. Consider two terms $t_1$ and $t_2$. We use JSD to assess the relevance of these terms to class $c_j$. Table 7 shows the calculation of JSD for the two terms. The total number of occurrences of all terms in $c_j$ is 100. The number of occurrences of $t_1$ and $t_2$ in $c_j$ is, respectively, 1 and 20. The reference collection of $c_j$, $c_{\text{ref}}$, has a total number of occurrences of all terms of 1000. The number of occurrences of $t_1$ and $t_2$ in $c_{\text{ref}}$ is, respectively, 225 and 10.

*Table 7: JSD calculation for two terms in a hypothetical case.*

|              | $t_1$            | $t_2$           |
|--------------|------------------|-----------------|
| $P(t)$       | 0.225 (225/1000) | 0.01 (10/1000)  |
| $Q(t)$       | 0.01 (1/100)     | 0.2 (20/100)    |
| $M(t)$       | 0.1175           | 0.105           |
| $JSD(c_j, t)$| 0.122            | 0.105           |

As shown in Table 7, $t_1$ has a higher JSD than $t_2$. This means that according to the JSD approach $t_1$ is a better label for $c_j$ than $t_2$. However, this is clearly a counter-intuitive result. To label $c_j$, we would expect $t_2$ to be preferred over $t_1$ because $t_2$ has both a higher frequency and a higher specificity. ∎

To avoid the problem shown in the above example, we consider two modifications of JSD. In the first modification we add the condition that $Q(t_i)$ must be greater than $P(t_i)$. This gives

$$JSD(t_i, c_j) = \begin{cases} P(t_i)\log\frac{P(t_i)}{M(t_i)} + Q(t_i)\log\frac{Q(t_i)}{M(t_i)} & \text{if } Q(t_i) > P(t_i) \\ 0 & \text{if } Q(t_i) \leq P(t_i) \end{cases} \quad (3)$$

As an alternative, we set the value of the first term in Eq. (2) to zero. This modified approach, which we denote as JSDQ, is defined by

$$JSDQ(t_i, c_j) = Q(t_i)\log\frac{Q(t_i)}{M(t_i)} \quad (4)$$

### *TF-IDF*

The TF-IDF approach is commonly used in the labeling literature (Mao et al., 2012; Muhr et al., 2010; Pourvali et al., 2019; Treeratpituk & Callan, 2006). This approach has been developed for quantifying the importance of terms in publications for information retrieval purposes (Salton & Buckley, 1988). TF-IDF combines the frequency of a term in a class with the log-transformed total frequency of the term in the reference collection. Like in the other approaches, we use the parent class as the reference collection and only take into account whether a term occurs in a publication or not (which differs from how TF-IDF is typically calculated in full-text information retrieval). TF-IDF is given by

$$\text{TF-IDF}(t_i, c_j) = tf(t_i, c_j) \times \log\frac{|c_p|}{tf(t_i, c_p)} \quad (5)$$

### *Waltman and van Eck approach*

Waltman and van Eck (2012) proposed an approach for term weighting that compares the frequency of $t_i$ in $c_j$ with the frequency of $t_i$ in $c_p$. We denote this approach as the WvE approach. The WvE approach includes a parameter $m$ that can be used to make a trade-off between frequency and specificity. The WvE approach is given by

$$WvE_m(t_i, c_j) = \frac{tf(t_i, c_j)}{tf(t_i, c_p) + m} \quad (6)$$

When $m$ is set to zero, WvE equals the frequency of $t_i$ in $c_j$ relative to the frequency of $t_i$ in $c_p$, which means that WvE is entirely focused on the specificity of terms. In contrast, when $m$ goes to infinity, WvE ranks terms according to their frequency. We test the WvE approach for the following values of $m$: 10, 25, 50, 75, 100, 1,000, 10,000 and 100,000.

*Term Frequency to Specificity*

In addition to term weighting approaches previously introduced in the literature and discussed above, we propose and test a new approach. Like WvE, this approach balances the frequency and specificity of a term in a class. The idea is to balance these two features in a more easily understandable way.

We denote the proposed approach as TFS (term frequency to specificity ratio). TFS is defined as

$$TFS_\alpha(t_i, c_j) = (ptf(t_i, c_j))^\alpha \times (s(t_i, c_j))^{1-\alpha} \qquad (7)$$

where the parameter $\alpha$ ($0 \leq \alpha \leq 1$) is used to weight frequency relative to specificity. $ptf(t_i, c_j)$ is the frequency of $t_i$ in $c_j$ relative to the number of publications in $c_j$. It is given by

$$ptf(t_i, c_j) = \frac{tf(t_i, c_j)}{|c_j|} \qquad (8)$$

The specificity of $t_i$ in $c_j$ is given by

$$s(t_i, c_j) = \frac{tf(t_i, c_j)}{expected} \qquad (9)$$

where the expected frequency of $t_i$ in $c_j$ is the same as for Chi-square (Eq. (1)).

TFS is the weighted geometric mean of *ptf* and *s*. In the analyses presented in this paper, the following values for the parameter $\alpha$ are considered: 0, 1/3, 1/2, 2/3 and 1. When $\alpha$ is set to 1/2, frequency and specificity have equal weight. When $\alpha$ is set to 1/3 (2/3), specificity (frequency) has more weight than frequency (specificity). Finally, when $\alpha$ is set to 0, TFS reduces to specificity, and likewise, when $\alpha$ is set to 1, TFS reduces to frequency.

**Evaluation methodology**

We use the Match@N rate for the evaluation (Carmel et al., 2009; Mao et al., 2012; Treeratpituk & Callan, 2006). For a given term weighting approach and bibliographic field, this measure indicates if any of the top $N$ ranked terms matches the baseline class label.

We first focus on the MeSH baseline. Consider a MeSH baseline class $c_l$. This class consists of all publications to which MeSH term $l$ has been assigned. For a given term weighting approach $A \in \{\text{Chi-square, JSD, JSDQ, TF-IDF, WvE, TFS}\}$, $A(t_i, c_l)$ is the relevance of term $t_i$ to $c_l$ according to $A$. We calculate $A(t_i, c_l)$ for each term $t_i$ occurring in at least three publications in $c_l$. Term $t_i$ is considered to occur in a publication if it occurs in one of the bibliographic fields used in the calculation of $A(t_i, c_l)$. Next, we rank the terms in descending order of $A(t_i, c_l)$. If one of the top $N$ most highly ranked terms matches the MeSH term $l$, the labeling is considered successful. Otherwise it is considered unsuccessful.

We now turn to the SMJC baseline. An SMJC category $l$ may have more than one label. For example, the category "Nanoscience & Nanotechnology" has two labels, "Nanoscience" and "Nanotechnology". The set of labels of $l$ is denoted by $S_l$. SMJC baseline class $c_l$ consists of all publications that have appeared in journals belonging to SMJC category $l$. For a given term weighting approach $A$, $A(t_i, c_l)$ is the relevance of term $t_i$ to $c_l$ according to $A$. We calculate $A(t_i, c_l)$ for each term $t_i$ occurring in at least three publications in $c_l$. Like in the case of the MeSH baseline, $t_i$ is considered to occur in a publication if it occurs in one of the bibliographic fields used in the calculation of $A(t_i, c_l)$. Terms are again ranked in descending order of $A(t_i, c_l)$. If one of the top $N$ most highly ranked terms matches a label in $S_l$, the labeling is considered successful. Otherwise it is considered unsuccessful. For example, in the case of the SMJC category "Nanoscience & Nanotechnology", the labeling is successful if either "Nanoscience" or "Nanotechnology" is included among the top $N$ terms.

Our evaluation measure is given by the proportion of classes that have been labeled successfully. The Match@N rate of term weighting approach $A$ relative to baseline $B \in \{\text{MeSH}, \text{SMJC}\}$, denoted by $\text{Match@N}(A, B)$, equals

$$\text{Match@N}(A, B) = \frac{n\_successful}{n\_total} \qquad (10)$$

where $n\_successful$ is the number of classes for which the labeling has been successful and $n\_total$ is the total number of classes.

For both the MeSH baseline and SMJC baseline, we set $N$ to 3 when calculating Match@N.[6] In our experience, including multiple terms in a label makes the label more descriptive and comprehensive. However, it also makes the interpretation of the class labels more time consuming for the user. By using three terms we try to balance these two aspects.

Some baseline labels do not occur at all in the publications belonging to the corresponding baseline class. Clearly, successful labeling is impossible in such a situation. In the results section, we therefore report the maximal possible Match@N rate for each combination of a baseline and bibliographic fields. The maximal possible Match@N rate is given by:

$$Max.possible = \frac{n\_extracted}{n\_total} \qquad (11)$$

where $n\_extracted$ is the number of classes for which a noun phrase corresponding to the class label has been extracted from the bibliographic field(s) in at least three publications, and $n\_total$ is the total number of classes.

---

[6] We also tested setting $N$ to 1 and 10. However, changing the value of $N$ did not lead to substantially different results. For this reason, we present the results only for $N = 3$.

# Results

In this section we report the results of our empirical analysis. First we report the results for the MeSH baseline, whereas the results for SMJC are given thereafter.

**MeSH baseline**

Figure 3 shows the Match@N rate for each combination of a term weighting approach and one or more bibliographic fields. The bar labeled "Max. possible" shows the proportion of MeSH classes for which the label could be extracted from the bibliographic fields. Depending on the bibliographic fields that are considered, this is the case for 60 to 80% of the MeSH classes. For the remaining classes the label cannot be found in the bibliographic fields (or it can be found in the bibliographic fields of only one or two publications). Figure 3 also shows the Match@N rate when terms are ranked based only on their frequency (denoted by $\text{TFS}_{\alpha=1}$; corresponding to the WvE approach with $m$ going to infinity) or only on their specificity (denoted by $\text{TFS}_{\alpha=0}$; corresponding to the WvE approach with $m = 0$).

*Figure 3: Match@N rates of different combinations of a term weighting approach and one or more bibliographic fields. Match@N rates are based on the MeSH baseline. Approaches are ranked in descending order of their Match@N rate obtained using titles and keywords.*

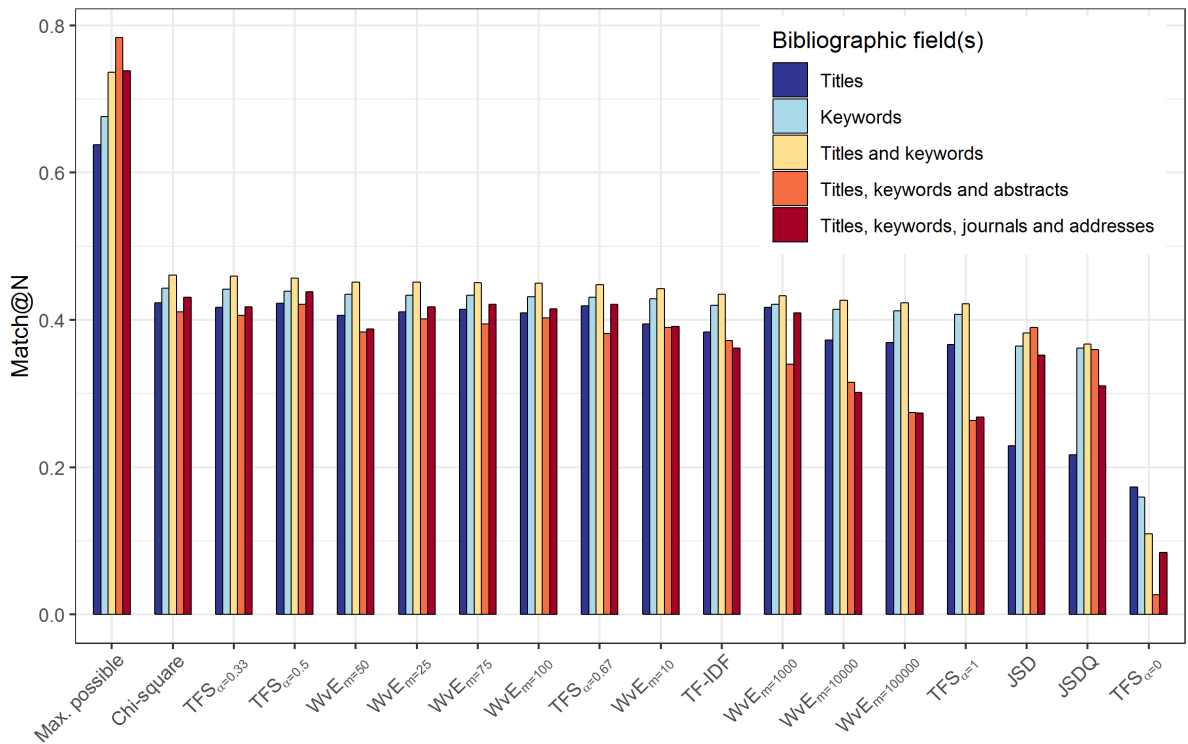

*Figure 4: Match@N rates of term weighting approaches using titles and keywords with 95% confidence intervals.* [7] *Match@N rates are based on the MeSH baseline.*

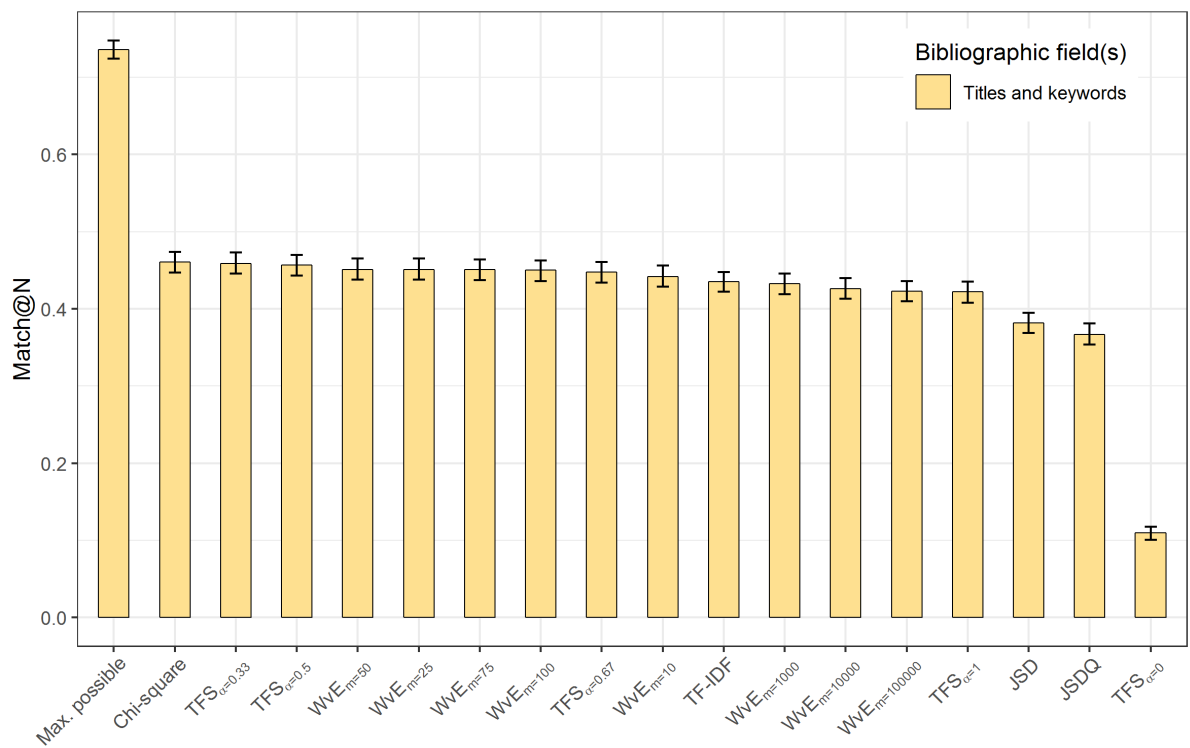

For almost all term weighting approaches, the combined use of titles and keywords yields the highest Match@N rate. This Match@N rate is higher than the Match@N rate obtained when only titles or only keywords are used. It is also higher than the Match@N rate obtained when titles and keywords are used in combination with abstracts or in combination with journals and addresses. The maximum possible Match@N rate is highest when titles and keywords are combined with abstracts, but in practice most term weighting approaches do not benefit from the use of abstracts.

Focusing on the results obtained using titles and keywords (Figure 4), Chi-square has the highest Match@N rate, but the difference with several other approaches is very small and the confidence intervals are overlapping. JSD, JSDQ and $TFS_{\alpha=0}$ have the lowest Match@N rate. WvE performs best with a relatively low value for the parameter $m$ (between 25 and 100). TFS performs best with values for the parameter $\alpha$ of 1/3 or 1/2, indicating that frequency and specificity should be given equal weight or that specificity should have somewhat more weight than frequency. However, giving full weight to specificity ($\alpha = 0$) leads to very bad results, while the results obtained by giving full weight to frequency ($\alpha = 1$) are more acceptable.

Table 8 shows some examples of the top 3-ranked terms by the Chi-square approach for "Bone Diseases" and "Vascular Diseases" and some child classes. If the MeSH term is included among the top 3-ranked terms, the relevant term is underlined in the table.

---

[7] Confidence intervals were obtained using the "binom.bayes()" function from the R package "binom".

Table 8: Top-ranked terms by Chi-square for some example classes within "Bone Diseases" and "Vascular Diseases" using the title and keywords fields. Terms that correspond to the class label are underlined. The MeSH number indicates the hierarchical structure of MeSH.

| Level | MeSH number | MeSH term | Top 3-ranked terms by Chi-square (ordered by rank) |
| --- | --- | --- | --- |
| 2 | C05.116 | Bone Diseases | osteoporosis; bone; bone mineral density |
| 3 | C05.116.900 | Spinal Diseases | spine; psoriatic arthritis; spondylitis |
| 4 | C05.116.900.153 | Intervertebral Disc Degeneration | disc degeneration; intervertebral disc; degenerative disc disease |
| 4 | C05.116.900.800 | Spinal Curvatures | idiopathic scoliosis; scoliosis; adolescent idiopathic scoliosis |
| 5 | C05.116.900.800.500 | Kyphosis | kyphoscoliosis; kyphoplasty; _kyphosis_ |
| 5 | C05.116.900.800.750 | Lordosis | _lordosis_; lumbar hyperlordosis; lumbar lordosis |
| 5 | C05.116.900.800.875 | Scoliosis | idiopathic scoliosis; scoliosis; adolescent idiopathic _scoliosis_ |
| 2 | C14.907 | Vascular Diseases | atherosclerosis; hypertension; stroke |
| 3 | C14.907.055 | Aneurysm | abdominal aortic aneurysm; aneurysm; intracranial _aneurysm_ |
| 4 | C14.907.055.625 | Iliac Aneurysm | iliac artery aneurysm; _iliac aneurysm_; aortoiliac aneurysm |
| 3 | C14.907.077 | Angiomatosis | sturge weber syndrome; klippel trenaunay syndrome; von hippel lindau disease |
| 4 | C14.907.077.410 | Klippel-Trenaunay-Weber Syndrome | _klippel trenaunay weber syndrome_; klippel trenaunay syndrome; venous malformation |
| 3 | C14.907.137 | Arterial Occlusive Diseases | atherosclerosis; carotid artery; coronary artery disease |
| 4 | C14.907.137.126 | Arteriosclerosis | atherosclerosis; coronary artery disease; inflammation |
| 5 | C14.907.137.126.307 | Atherosclerosis | _atherosclerosis_; inflammation; macrophage |

Using titles and keywords, Figure 5 shows the Match@N rates of different term weighting approaches for MeSH classes at different levels in the MeSH tree. The results obtained for the various approaches show a quite consistent pattern. Ignoring levels 11 to 13, for which the number of MeSH classes is very small, we observe the highest Match@N rates at levels 5 and 6, indicating that the different approaches perform best for MeSH classes that are neither very broad nor very specific.

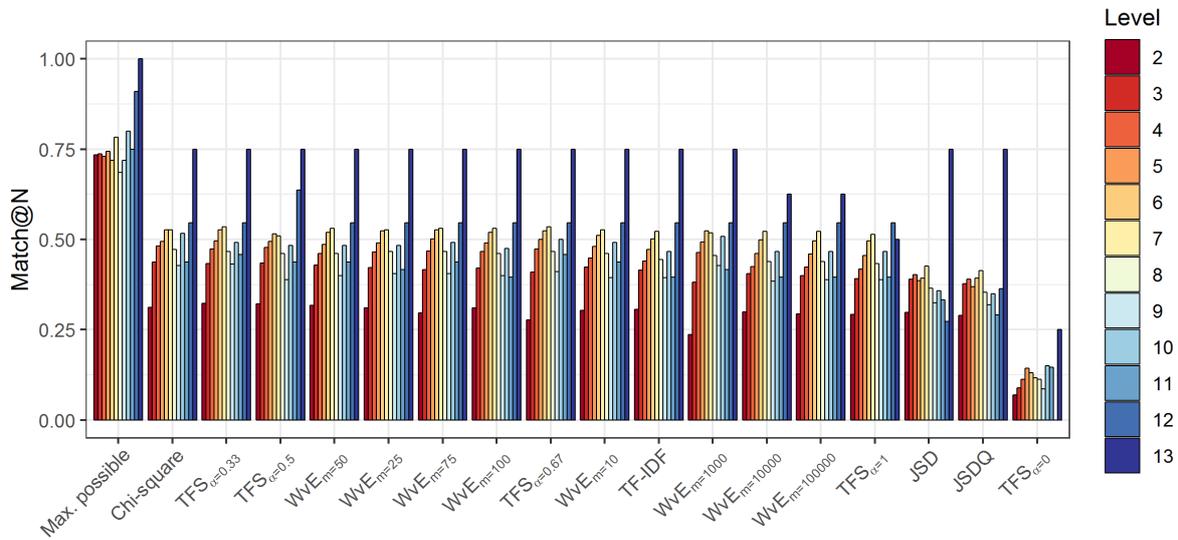

*Figure 5: Match@N rates of different term weighting approaches for MeSH classes at different levels in the MeSH tree. Match@N rates were obtained using titles and keywords.*

**SMJC baseline**

As shown in Figure 6, for a majority of the approaches the highest Match@N rate is obtained by the combined use of four bibliographic fields: journals, addresses, titles and keywords. However, the Match@N rates obtained using journals and addresses without titles and keywords are almost as high. The use of only addresses tends to yield a somewhat lower Match@N rate, while the use of only journals gives an even lower Match@N rate. The lowest Match@N rate is obtained by using titles and keywords without journals and addresses. Hence, compared to titles and keywords, journals and addresses provide a vocabulary that better resembles disciplinary labels such as those used in the SMJC.

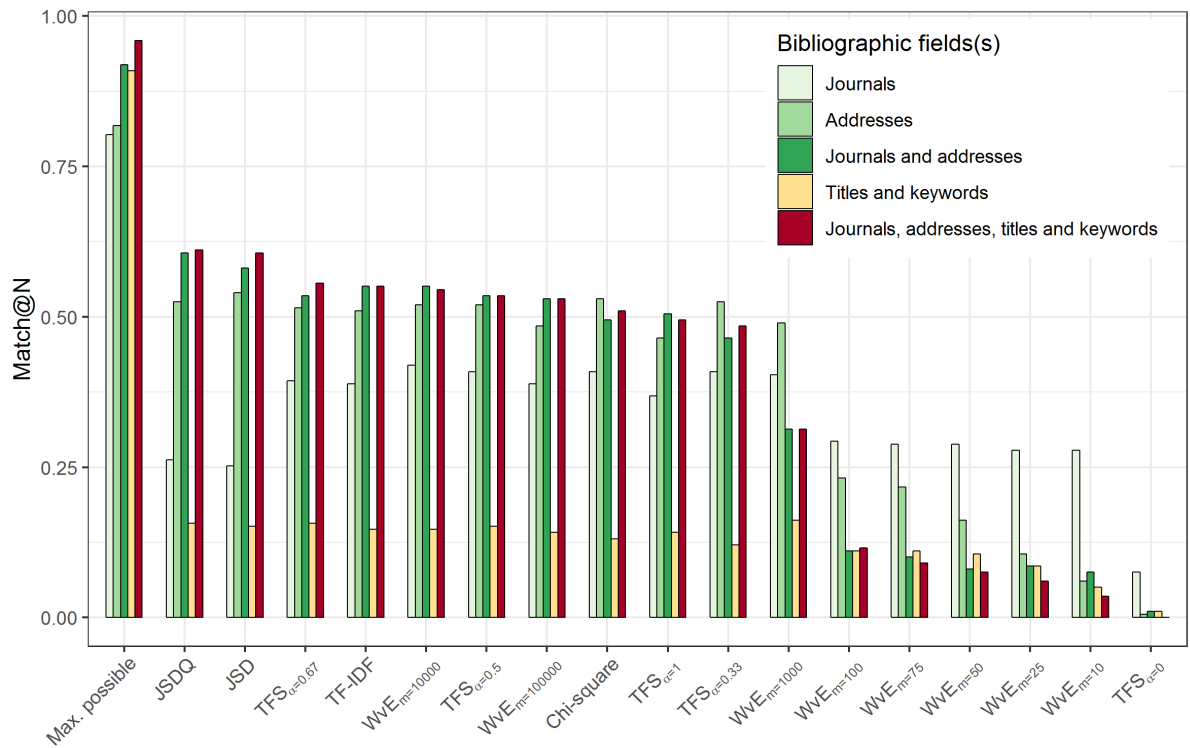

*Figure 6: Match@N rates of different combinations of a term weighting approach and one or more bibliographic fields. Match@N rates are based on the SMJC baseline. Approaches are ranked in descending order of their Match@N rate obtained using journals, addresses, titles and keywords.*

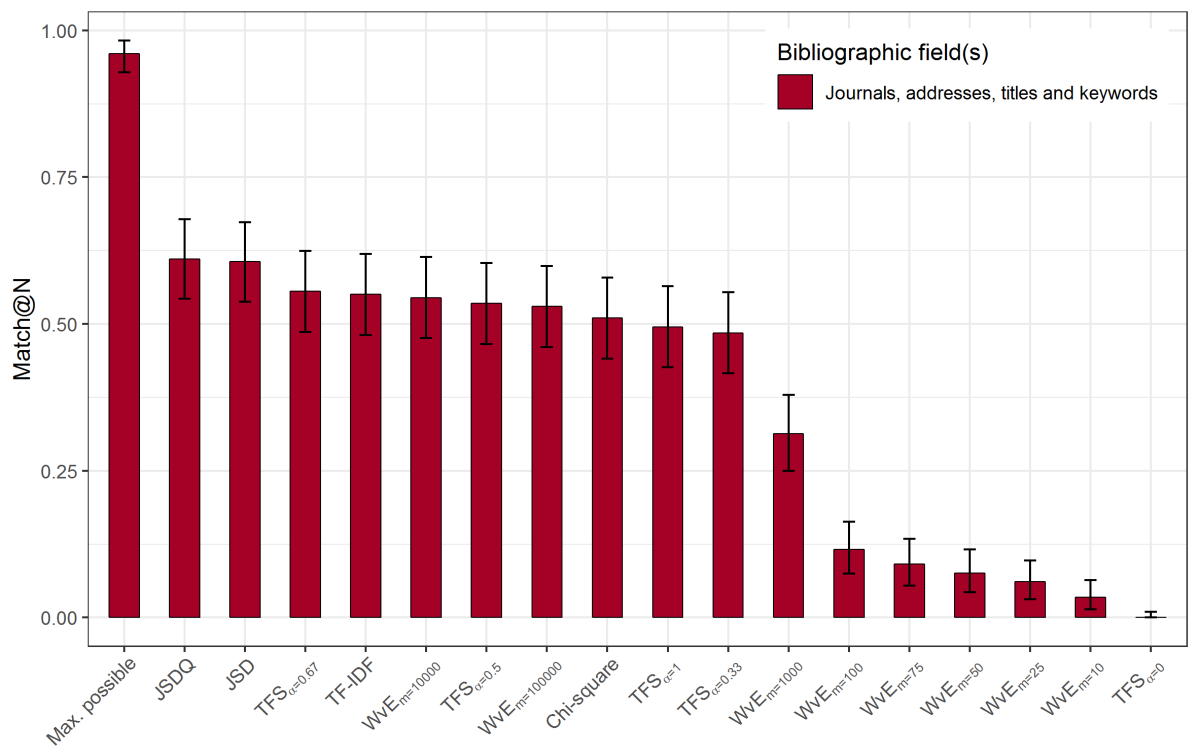

*Figure 7: Match@N rates of term weighting approaches using journals, addresses, titles and keywords with 95% confidence intervals. Match@N rates are based on the SMJC baseline.*

Interestingly, the JSD and JSDQ approaches perform much better using the SMJC baseline than using the MeSH baseline. JSD and JSDQ outperform all other approaches, although in many cases the differences in Match@N rate are not very large. Figure 7 shows the results with 95% confidence intervals for the fields journals, addresses, titles and keywords. JSDQ is the best performing approach, but its confidence interval is overlapping with the confidence intervals of many of the other approaches. TFS performs better for higher values of the parameter $\alpha$ and likewise WvE performs better for higher values of the parameter *m*. This shows that for labeling of SMJC classes frequency should be given a relatively high weight, while specificity should have a relatively low weight. This is different from labeling of MeSH classes, where we have seen that frequency should have less weight.

Table 9 shows some example classes and top 3-ranked terms by JSDQ using the journals, addresses, titles and keywords fields. Terms that correspond to the SMJC class label are underlined.

*Table 9: Top-ranked terms by JSDQ for some example classes in SMJC using the journals, addresses, titles and keywords fields. Terms that correspond to the class label are underlined.*

| Level | SMJC label | Top 3-ranked terms by JSDQ (ordered by rank) |
|---|---|---|
| 3 | Analytical Chemistry | analytical chemistry; determination; mass spectrometry |
| 3 | Biochemistry & Molecular Biology | biochemical; biochemistry; biological chemistry |
| 3 | Dentistry | dentistry; journal; oral |
| 3 | Microbiology | infection; infectious disease; microbiology |
| 3 | Social Psychology | personality; social psychology; psychology |
| 3 | Speech-Language Pathology & Audiology | language; speech; language processing |
| 3 | Urban & Regional Planning | geography; planning; urban planning |
| 3 | International Relations | international study; polit science; international affairs |
| 3 | Civil Engineering | civil; civil engineering; structural engineering |
| 2 | Clinical Medicine | patient; surgery; medicine |

## Discussion and conclusions

Using two baseline classifications, one based on MeSH and one based on SMJC, we have evaluated different approaches to label the classes in an algorithmically constructed publication-level classification (ACPLC). We have focused on two important choices: the choice of (1) different bibliographic fields and (2) different approaches to weight the relevance of terms. Terms were extracted from titles, abstracts, author keywords, journal names and author addresses. The relevance of terms was determined using different term weighting approaches. The results were evaluated using the two baselines. Based on our findings, we present three recommendations.

Our results show that the level of aggregation needs to be taken into account when labeling the classes in an ACPLC. Terms extracted from titles and keywords are more similar to MeSH terms than terms extracted from other bibliographic fields. The performance of our labeling approaches is especially high for classes at levels 4 to 7 in the MeSH tree. In general these classes represent topics at a rather detailed level. Terms extracted from addresses and journals express subject orientation at higher semantic levels and the results using the MeSH baseline suggest that such terms are not suitable to label classes corresponding to narrow research topics. Abstracts on the other hand can be expected to contain terms at lower semantic levels. Nevertheless, our results indicate that using abstracts decreases the quality of the class labels, probably because abstracts contain terms that are less focused in comparison with titles and keywords.

Recommendation 1:
➢ Extract terms from titles and keywords to label classes at the topic level.

SMJC categories represent relatively broad disciplines. Terms extracted from journals and addresses are more similar to the labels of SMJC categories than terms extracted from titles and keywords. The use of titles and keywords in addition to journals and addresses has almost no added value.

Recommendation 2:
➢ Extract terms from journals and addresses to label classes at the discipline level.

When we compare the results for the MeSH baseline with the results for the SMJC baseline, the relative performance of the term weighting approaches is almost the opposite. We therefore conclude that the balance between term specificity and term frequency should be adjusted to the granularity level. JSDQ, JSD and TF-IDF all give high weight to term frequency relative to term specificity. For this reason, these approaches perform well in comparison with other approaches at high semantic levels. By contrast, Chi-square favors high specificity and performs well at lower semantic levels. For other baselines not considered in this paper, we also expect the performance of the term weighting approaches to depend on the semantic level of the classes in the baseline.

The WvE and TFS approaches both include a parameter that can be adjusted. The two approaches both interpolate between frequency and specificity and use a parameter to adjust the balance. When the MeSH baseline is used for evaluation, the results show that both approaches perform better when the respective parameter values are low (i.e. when specificity is favored). On the other hand, when the SMJC baseline is used, both WvE and TFS perform better when the respective parameter values are high (i.e. when frequency is favored).

The parameters of the WvE and TFS approaches make it possible to obtain good results at different semantic levels. In our view the interpolation between frequency and specificity can be more easily understood in TFS, since frequency and specificity are clearly separated in the definition of TFS (Eq. (7)). In addition, the parameter $\alpha$ of TFS has a more intuitive meaning than the parameter $m$ of WvE. Our results also show that TFS performs relatively well at different levels of granularity using the same parameter value (0.5). WvE does not perform as well using a fixed parameter value.

Recommendation 3:
➢ Use TFS to calculate term relevance for labeling classes in a hierarchical ACPLC. Set the parameter $\alpha$ to 0.5 as default. Adjust $\alpha$ to higher values for classes at lower levels of granularity and to lower values for classes at higher levels of granularity.

This study has focused on the use of different bibliographic fields and term-weighting approaches for labeling of classes. In future studies it is of interest to also take into account different term extraction approaches. The term extraction approach used in this study considers only labels obtained from phrases present in the bibliographic fields of publications. The examples in Table 8 indicate that the methodology is more effective in labeling specific topics, such as specific diseases (e.g. Kyphosis, Lordosis, Scoliosis), than in creating labels at a more general level (e.g. Spinal Curvatures). This issue needs further exploration. Approaches that use external sources such as Wikipedia or thesauri may be more successful in creating more general labels. Thesauri also provide information about term relations and can be used to merge synonyms.

Our results indicate that abstracts add more noise than they add value. Nevertheless, abstracts can be expected to contain valuable information. More advanced techniques for term extraction and relevance ranking may be able to use this information to obtain better labels. Dependency parsing (Z. Li et al., 2015) and Word2vec (Mikolov et al., 2013) are examples of approaches that may be helpful.

In this study we have given equal weight to terms in different bibliographic fields. Future studies may explore whether giving different weights could improve class labeling. Another approach that may improve the labeling of classes is suggested by Mao et al. (2012). For classes that have children, this approach takes into account the way a term is distributed over the child classes. Terms that are more equally distributed over the child classes are preferred over terms that are unevenly distributed. The approach of Mao et al. would require further development in order to be used to label classes in a hierarchical ACPLC.

Other improvements may be achieved using stop lists. We have seen that some classes are labeled with terms such as "journal", "effect" and "result", which are clearly of low relevance (see for example the top 3-ranked terms obtained for "Dentistry" in Table 9). Such terms could be removed using a stop list.

There are several issues that we have not been able to address in this study. For example, the MeSH baseline focuses on biomedical and health research. It is not clear to what extent our results generalize to other research areas, which are characterized by different vocabulary, different styles for titles and abstracts and different types of author keywords. Furthermore, we know little about the way labels are perceived by users. In-depth user studies are needed to get a detailed understanding of the way users experience the strengths and weaknesses of different labeling approaches.

## Acknowledgments

We would like to thank four anonymous reviewers for their relevant and constructive comments on an earlier version of this paper.

## Funding information

Peter Sjögårde was funded by The Foundation for Promotion and Development of Research at Karolinska Institutet.

## Competing interests

Ludo Waltman is affiliated with the Centre for Science and Technology Studies (CWTS) at Leiden University. CWTS uses classifications similar to the ones discussed in this paper in commercial applications.

## References

Aggarwal, C. C., & Zhai, C. (2012). A survey of text clustering algorithms. In C. C. Aggarwal & C. Zhai (Eds.), *Mining Text Data* (pp. 77–128). Springer US. https://doi.org/10.1007/978-1-4614-3223-4_4

Ahlgren, P., Chen, Y., Colliander, C., & van Eck, N. J. (2020). Enhancing direct citations: A comparison of relatedness measures for community detection in a large set of PubMed publications. *Quantitative Science Studies*, 1–17. https://doi.org/10.1162/qss_a_00027

Ahlgren, P., Colliander, C., & Sjögårde, P. (2018). Exploring the relation between referencing practices and citation impact: A large-scale study based on Web of Science data. *Journal of the Association for Information Science and Technology*, *69*(5), 728–743. https://doi.org/10.1002/asi.23986

Allahyari, M., & Kochut, K. (2015). Automatic topic labeling using ontology-based topic models. *2015 IEEE 14th International Conference on Machine Learning and Applications (ICMLA)*, 259–264. https://doi.org/10.1109/ICMLA.2015.88

Archambault, É., Beauchesne, O., & Caruso, J. (2011). Towards a multilingual, comprehensive and open scientific journal ontology. *Proceedings of the 13th International Conference on Scientometrics and Infometrics, Durban, South Africa*, 66–77. http://www.science-metrix.com/pdf/Towards_a_Multilingual_Comprehensive_and_Open.pdf

Blei, D. M., & Lafferty, J. D. (2007). A correlated topic model of science. *The Annals of Applied Statistics*, *1*(1), 17–35.

Blei, D. M., & Lafferty, J. D. (2009). Topic models. In A. Srivastava & M. Sahami, *Text Mining: Theory and applications* (pp. 71–94). Taylor and Francis. https://doi.org/10.1201/9781420059458-12

Boyack, K. W., & Klavans, R. (2014). Including cited non-source items in a large-scale map of science: What difference does it make? *Journal of Informetrics*, *8*(3), 569–580. https://doi.org/10.1016/j.joi.2014.04.001

Boyack, K. W., & Klavans, R. (2020). A comparison of large-scale science models based on textual, direct citation and hybrid relatedness. *Quantitative Science Studies*, 1–19. https://doi.org/10.1162/qss_a_00085

Carmel, D., Roitman, H., & Zwerdling, N. (2009). Enhancing cluster labeling using Wikipedia. *Proceedings of the 32Nd International ACM SIGIR Conference on Research and Development in Information Retrieval*, 139–146. https://doi.org/10.1145/1571941.1571967

Carmel, D., Yom-Tov, E., Darlow, A., & Pelleg, D. (2006). What makes a query difficult? *Proceedings of the 29th Annual International ACM SIGIR Conference on Research and Development in Information Retrieval*, 390–397. https://doi.org/10.1145/1148170.1148238

Cutting, D. R., Karger, D. R., Pedersen, J. O., & Tukey, J. W. (1992). Scatter/gather: A cluster-based approach to browsing large document collections. *Proceedings of the 15th Annual International ACM SIGIR Conference on Research and Development in Information Retrieval*, 318–329. https://doi.org/10.1145/133160.133214

Deerwester, S., Dumais, S. T., Furnas, G. W., Landauer, T. K., & Harshman, R. (1990). Indexing by latent semantic analysis. *Journal of the American Society for Information Science*, *41*(6), 391–407. https://doi.org/10.1002/(SICI)1097-4571(199009)41:6<391::AID-ASI1>3.0.CO;2-9

Hennig, P., Berger, P., Steuer, C., Wuerz, C., & Meinel, C. (2014). Cluster labeling for the blogosphere. *2014 IEEE Fourth International Conference on Big Data and Cloud Computing*, 416–423. https://doi.org/10.1109/BDCloud.2014.68

Hotho, A., Staab, S., & Stumme, G. (2003). Ontologies improve text document clustering. *Third IEEE International Conference on Data Mining*, 541–544. https://doi.org/10.1109/ICDM.2003.1250972

Justeson, J. S., & Katz, S. M. (1995). Technical terminology: Some linguistic properties and an algorithm for identification in text. *Natural Language Engineering*, *1*(1), 9–27. https://doi.org/10.1017/S1351324900000048

Kageura, K., & Umino, B. (1996). Methods of automatic term recognition: A review. *Terminology*, *3*(2), 259–289.

Klavans, R., & Boyack, K. W. (2017). Which type of citation analysis generates the most accurate taxonomy of scientific and technical knowledge? *Journal of the Association for Information Science and Technology*, *68*(4), 984–998. https://doi.org/10.1002/asi.23734

Koopman, R., Wang, S., & Scharnhorst, A. (2017). Contextualization of topics: Browsing through the universe of bibliographic information. *Scientometrics*, *111*(2), 1119–1139. https://doi.org/10.1007/s11192-017-2303-4

Li, S.-T., & Tsai, F.-C. (2010). Constructing tree-based knowledge structures from text corpus. *Applied Intelligence*, *33*(1), 67–78. https://doi.org/10.1007/s10489-010-0243-2

Li, Z., Li, J., Liao, Y., Wen, S., & Tang, J. (2015). Labeling clusters from both linguistic and statistical perspectives: A hybrid approach. *Knowledge-Based Systems*, *76*, 219–227. https://doi.org/10.1016/j.knosys.2014.12.019

Manning, C. D., Raghavan, P., & Schütze, H. (2008). *Introduction to information retrieval*. Cambridge University Press.

Manning, C. D., Surdeanu, M., Bauer, J., Finkel, J. R., Bethard, S., & McClosky, D. (2014). *The Stanford CoreNLP natural language processing toolkit*. 55–60. https://doi.org/10.3115/v1/P14-5010

Mao, X.-L., Ming, Z.-Y., Zha, Z.-J., Chua, T.-S., Yan, H., & Li, X. (2012). Automatic labeling hierarchical topics. *Proceedings of the 21st ACM International Conference on Information and Knowledge Management - CIKM '12*, 2383–2386. https://doi.org/10.1145/2396761.2398646

Mikolov, T., Chen, K., Corrado, G., & Dean, J. (2013). Efficient estimation of word representations in vector space. *ArXiv:1301.3781 [Cs]*. http://arxiv.org/abs/1301.3781

Milanez, D. H., Noyons, E., & de Faria, L. I. L. (2016). A delineating procedure to retrieve relevant publication data in research areas: The case of nanocellulose. *Scientometrics*, *107*(2), 627–643. https://doi.org/10.1007/s11192-016-1922-5

Muhr, M., Kern, R., & Granitzer, M. (2010). Analysis of structural relationships for hierarchical cluster labeling. *Proceedings of the 33rd International ACM SIGIR Conference on Research and Development in Information Retrieval*, 178–185. https://doi.org/10.1145/1835449.1835481

*NLM Medical Text Indexer (MTI)*. (n.d.). Retrieved 5 December 2019, from https://ii.nlm.nih.gov/MTI/index.shtml

Perianes-Rodriguez, A., & Ruiz-Castillo, J. (2017). A comparison of the Web of Science and publication-level classification systems of science. *Journal of Informetrics*, *11*(1), 32–45. https://doi.org/10.1016/j.joi.2016.10.007

Pourvali, M., Orlando, S., & Omidvarborna, H. (2019). Topic Models and Fusion Methods: A Union to Improve Text Clustering and Cluster Labeling. *International Journal of Interactive Multimedia and Artificial Intelligence*, *5*(4), 28–34. https://doi.org/10.9781/ijimai.2018.12.007

Ruiz-Castillo, J., & Waltman, L. (2015). Field-normalized citation impact indicators using algorithmically constructed classification systems of science. *Journal of Informetrics*, *9*(1), 102–117. https://doi.org/10.1016/j.joi.2014.11.010

Salton, G., & Buckley, C. (1988). Term-weighting approaches in automatic text retrieval. *Information Processing & Management*, *24*(5), 513–523. https://doi.org/10.1016/0306-4573(88)90021-0

Seifert, C., Sabol, V., Kienreich, W., Lex, E., & Granitzer, M. (2014). Visual analysis and knowledge discovery for text. In A. Gkoulalas-Divanis & A. Labbi (Eds.), *Large-Scale Data Analytics* (pp. 189–218). Springer New York. https://doi.org/10.1007/978-1-4614-9242-9_7

Sjögårde, P., & Ahlgren, P. (2018). Granularity of algorithmically constructed publication-level classifications of research publications: Identification of topics. *Journal of Informetrics*, *12*(1), 133–152. https://doi.org/10.1016/j.joi.2017.12.006

Sjögårde, P., & Ahlgren, P. (2020). Granularity of algorithmically constructed publication-level classifications of research publications: Identification of specialties. *Quantitative Science Studies*, *1*(1), 207–238. https://doi.org/10.1162/qss_a_00004

Small, H., Boyack, K. W., & Klavans, R. (2014). Identifying emerging topics in science and technology. *Research Policy*, *43*(8), 1450–1467. https://doi.org/10.1016/j.respol.2014.02.005

Spärck Jones, K. (2004). A statistical interpretation of term specificity and its application in retrieval. *Journal of Documentation*, *60*(5), 493–502. https://doi.org/10.1108/00220410410560573

Šubelj, L., van Eck, N. J., & Waltman, L. (2016). Clustering scientific publications based on citation relations: A systematic comparison of different methods. *PLOS ONE*, *11*(4), e0154404. https://doi.org/10.1371/journal.pone.0154404

Suominen, A., & Toivanen, H. (2016). Map of science with topic modeling: Comparison of unsupervised learning and human-assigned subject classification. *Journal of the Association for Information Science and Technology*, *67*(10), 2464–2476. https://doi.org/10.1002/asi.23596

Toutanova, K., Klein, D., Manning, C. D., & Singer, Y. (2003). Feature-rich part-of-speech tagging with a cyclic dependency network. *Proceedings of the 2003 Conference of the North American Chapter of the Association for Computational Linguistics on Human Language Technology - Volume 1*, 173–180. https://doi.org/10.3115/1073445.1073478

Toutanova, K., & Manning, C. D. (2000). Enriching the knowledge sources used in a maximum entropy part-of-speech tagger. *Proceedings of the 2000 Joint SIGDAT Conference on Empirical Methods in Natural Language Processing and Very Large Corpora: Held in Conjunction with the 38th Annual Meeting of the Association for Computational Linguistics*, *13*, 63–70. https://doi.org/10.3115/1117794.1117802

Treeratpituk, P., & Callan, J. (2006). Automatically labeling hierarchical clusters. *Proceedings of the 2006 National Conference on Digital Government Research*, 167–176. https://doi.org/10.1145/1146598.1146650

Velden, T., Boyack, K. W., Gläser, J., Koopman, R., Scharnhorst, A., & Wang, S. (2017). Comparison of topic extraction approaches and their results. *Scientometrics*, *111*(2), 1169–1221. https://doi.org/10.1007/s11192-017-2306-1

Velden, T., Yan, S., & Lagoze, C. (2017). Mapping the cognitive structure of astrophysics by infomap clustering of the citation network and topic affinity analysis. *Scientometrics*, *111*(2), 1033–1051. https://doi.org/10.1007/s11192-017-2299-9

Waltman, L., Boyack, K. W., Colavizza, G., & van Eck, N. J. (2020). A principled methodology for comparing relatedness measures for clustering publications. *Quantitative Science Studies*, 1–36. https://doi.org/10.1162/qss_a_00035

Waltman, L., & van Eck, N. J. (2012). A new methodology for constructing a publication-level classification system of science. *Journal of the American Society for Information Science and Technology*, *63*(12), 2378–2392. https://doi.org/10.1002/asi.22748

Wang, Q., & Ahlgren, P. (2018). Measuring the interdisciplinarity of research topics. In *STI 2018 Conference Proceedings* (pp. 134–142). https://openaccess.leidenuniv.nl/handle/1887/65325